\documentclass[12pt]{iopart}

\begin{document}
\hspace*{3.5 in}CUQM-116\\
\hspace*{3.5 in}math-ph/0605057\\

\title[Sextic anharmonic oscillators and orthogonal polynomials]{Sextic anharmonic oscillators and orthogonal polynomials}
\author{Nasser Saad$^1$, Richard L. Hall$^2,$ and Hakan Ciftci$^3$}

\address{$^1$ Department of Mathematics and Statistics,
University of Prince Edward Island,
550 University Avenue, Charlottetown,
PEI, Canada C1A 4P3.}
\address{$^2$ Department of Mathematics and Statistics, Concordia University,
1455 de Maisonneuve Boulevard West, Montr\'eal,
Qu\'ebec, Canada H3G 1M8}
\address{$^3$ Gazi Universitesi, Fen-Edebiyat Fak\"ultesi, Fizik
B\"ol\"um\"u, 06500 Teknikokullar, Ankara, Turkey.}
\eads{\mailto{nsaad@upei.ca}, \mailto{rhall@mathstat.concordia.ca}, \mailto{hciftci@gazi.edu.tr}}

\begin{abstract}
Under certain constraints on the parameters $a$, $b$ and $c$, it is known that Schr\"odinger's equation $-{d^2\psi/dx^2}+(ax^6+bx^4+cx^2)\psi=E\psi,~a>0$ with the sextic anharmonic oscillator potential is exactly solvable. In this article we show that the exact wave function $\psi$ is the generating function for a set of orthogonal polynomials $\{P_n^{(t)}(x)\}$ in the energy variable $E$. Some of the properties of these polynomials are discussed in detail and our analysis reveals scaling and factorization properties that are central to quasi-exact solvability. We also prove that this set of orthogonal polynomials can be reduced, by means of a simple scaling transformation, to a remarkable class of orthogonal polynomials, $P_n(E)=P_n^{(0)}(E)$ recently discovered by Bender and Dunne.
\end{abstract}

\vskip0.2in


\maketitle

\section{Introduction: }
Recently, Bender and Dunne \cite{bd} introduced a remarkable set of orthogonal polynomials associated with the one-dimensional Hamiltonian
\begin{equation}\label{eq1}
H=-{d^2\over dx^2}+x^6-(4s+4J-2)x^2.
\end{equation}
where $J$ is a positive integer and $s=1/4$ or $s=3/4$. The two choices of $s$ correspond respectively to the even-parity and odd-parity solutions $\psi_E(x)$ of the eigen-equation $H\psi = E\psi.$  Bender and Dunne showed that $\psi_E(x)$ is the generating function for a set of orthogonal polynomials $\{P_n(E)\}$ in the energy variable $E$. These polynomials may easily be shown to satisfy the three-term recursion relation (with $P_0(E)=1$, $P_1(E)=E$)
\begin{equation}\label{eq2}
P_n(E)=EP_{n-1}(E)+16(n-1)(n-J-1)(n+2s-2)P_{n-2},\quad n\geq 2
\end{equation}
from which it follows that they are orthogonal with respect to a certain weight function $\omega(E)$:
\begin{equation}\label{eq3}
\int P_n(E)P_k(E)\omega(E)dE=0,\quad n\neq k
\end{equation}
The `weight function' $\omega(E)$ (which we note is not necessarily positive) can be constructed by an algebraic method discussed in detail in Ref. \cite{ak}.  If the initial conditions $P_0(E)=1$ and $P_1(E)=E,$
are imposed, each coefficient $P_n(E)$ becomes a monic polynomial of degree $n.$ The form of the coefficients of the recursion relation satisfied by the polynomial system $\{P_n(E)\}$ implies that this system has several remarkable properties. First, the squared norms of the polynomials $P_n(E)$ vanish for $n\geq J+1$ if $J$ is a positive integer. Secondly, each $P_n(E)$, with $n\geq J+1$, factors into a single product of $P_{J+1}$ with another polynomial, i.e. 
\begin{equation}\label{eq4}
P_{J+m+1}(E)=P_{J+1}(E)Q_m(E), ~~m\geq 0.
\end{equation}
These factor-polynomials $Q_m(E)$ form an orthogonal set. There are a number of articles devoted to the study the properties of these orthogonal polynomials \cite{ak}-\cite{mm}. The purpose of the present article is to study a set of orthogonal polynomials $\{P_n^{(t)}(E)\}$ associated with the one-dimensional sextic anharmonic oscillator Hamiltonian \cite{dw}-\cite{vs}
\begin{equation}\label{eq5}
H(a,b,c)=-{d^2\over dx^2}+ax^6+bx^4+cx^2,\quad a> 0,
\end{equation}
in which the potential's parameters obey certain constraints. We show that for certain constraints on $a,$ $b$ and $c$,  the wave function solution of the Hamiltonian $H(a,b,c)$ is the generating function for a set of orthogonal polynomials in the energy variable $E$. We explicity construct the polynomial solvability constraints and prove that they obey a three-term recursion relation; consequently, they form a set of orthogonal polynomials \cite{fj}. We study some of the properties of these polynomials such as: for nonnegative integer values of $J$, for which $H(a,b,c)$ is quasi-exactly solvable, the squared norms of the polynomials $P_n^{(t)}(E)$ vanish for $n\geq J+1$. Further, the polynomials $P_{n}^{(t)}(E)$ of degree higher than $J+1$ factor into a product of two polynomials, one of which is $P_{J+1}^{(t)}(E)$. We also show that under a scale transformation, they leads to the Bender-Dunne class of orthogonal polynomials $P_n(E)=P_n^{(0)}(E)$. To this end the paper is organized as follows. Section 2 contains a general technique for generating the polynomial solvability  constraints of the sextic anharmonic oscillator Hamiltonian (\ref{eq5}); these can be easily extended to study the exact solutions for Hamiltonians with even-degree polynomials. We further show, through an explicit construction, that the wave function solution is a generating function of these polynomials. We prove thereafter that this set of polynomials satisfy a three-term recursion relation, and consequently they form a class of orthogonal polynomials. In section 4, we show under simple scaling transformation the correspondence between a quasi-exact solvable model and a set of orthogonal polynomials, and we show that, under suitable change of variables, they generalize the class of Bender-Dunne orthogonal polynomials \cite{bd}.
In the appendix we show that some of these polynomials can be expressed in terms of Meixner polynomials of the second kind.
 While the results in the present work including constraining relations for the potential parameters, they can usefully be compared with the ${\mathcal P}{\mathcal T}$-symmetric version of complex sextic potentials, recently studied by Bender and Monou \cite{bm}, and the work of Bender and Turbiner \cite{bt}.
\section{Solvability constraints of the sextic anharmonic oscillator Hamiltonian}
Let us assume that the exact solution of the Schr\"odinger equation
\begin{equation}\label{eq6}
-\psi''(x)+V(x)\psi=E\psi
\end{equation}
takes the form
$\psi(x)=\chi(x)e^{-f(x)}$. On direct substitution in (\ref{eq6}), we obtain the following equation for $\chi(x)$
\begin{equation}\label{eq7}
\chi''(x)=2f'(x)\chi'(x)+(f''(x)-f'^2(x)+V(x)-E)\chi(x).
\end{equation}
Without loss of generality, we may assume, for the node-less eigenstate, that $\chi(x)$ is a constant, $\chi(x)=1$. In this case, Eq.(\ref{eq7}) reads
\begin{equation}\label{eq8}
u'(x)=E-V(x)+u^2(x),\quad\quad (u(x)=f'(x)),
\end{equation}
which is a special form of a Riccati equation. For the sextic anharmonic oscillator potential $V(x)=ax^6+bx^4+cx^2$, we can solve this differential equation exactly for certain constraints on the parameters $a$, $b$ and $c$. The solvability of this differential equation is based on an elegant approach introduced earlier by Rainville \cite{ra} providing necessary conditions for polynomial solutions of certain Riccati equations. 
\vskip0.1in
\noindent\emph{Definition 1: By the symbol $[\sqrt{P(x)}]$, where $P(x)$ is a polynomial of even-degree, we shall mean the polynomial part of the expansion of $\sqrt{P(x)}$ in a series of descending integral powers of $x$.} For example:
\begin{equation}\label{eq9}
\bigg[\sqrt{x^6-4x^4+7x^2-2}\bigg]=x^3-2x.
\end{equation}
\vskip0.1in
With this notation we may state (for a proof, see \cite{ra})
\vskip0.1in
\noindent\emph{Theorem 1: If in
\begin{equation}\label{eq10}
{du\over dx}=A_0(x)+u^2
\end{equation}
$A_0(x)$ is a polynomial of even-degree, then no polynomial other than
\begin{equation}\label{eq11}
u=\pm[\sqrt{-A_0}]
\end{equation}
can be a solution of (\ref{eq10}). If the degree of $A_0$ is odd, there is no polynomial solution of (\ref{eq10}).} 
As an example, for the first-order nonlinear differential equation ${u'}=2-7x^2+4x^4-x^6+u^2$, we have $u(x)=-(x^3-2x)$ is a solution which fact can be easily verify through direct substitution.
\vskip0.1in
\noindent By means of this theorem, we can search for exact solutions of the differential equation (\ref{eq8}) with $V(x)=ax^6+bx^4+cx^2$, in which case we have 
\begin{equation}\label{eq12}
u=\pm\bigg[\sqrt{ax^6+bx^4+cx^2-E}\bigg]=\pm(\sqrt{a}x^3+{b\over 2\sqrt{a}}x),
\end{equation}
if $12{a}^{3\over 2}-b^2+4ac=0$ and $E={b\over 2\sqrt{a}}$. For a physical acceptable solution satisfying $\psi(\pm\infty)=0$, we have, for $u(x)=f'(x)$, that
\begin{equation}\label{eq13}
f(x)={\sqrt{a}\over 4}x^4+{b\over 4\sqrt{a}}x^2
\end{equation}
Consequently, for the Schr\"odinger equation
\begin{equation}\label{eq14}
-\psi''(x)+(ax^6+bx^4+cx^2)\psi=E\psi,
\end{equation}
we may assume that exact solution takes the form
\begin{equation}\label{eq15}
\psi(x)=\chi(x)e^{-{\sqrt{a}\over 4}x^4-{b\over 4\sqrt{a}}x^2}
\end{equation}
which has been adopted in the literature for this class of potentials \cite{dw}-\cite{vs}. In order to find $\chi(x)$, we notice by means of  (\ref{eq15}) and (\ref{eq7}) that
\begin{equation}\label{eq16}
\chi''=2(\sqrt{a}x^3+{b\over 2\sqrt{a}}x)\chi'+({b\over 2\sqrt{a}}-E+(3\sqrt{a}-{b^2\over 4a}+c)x^2)\chi
\end{equation}
which clearly yields a ground-state eigenenergy $E={b\over 2\sqrt{a}}$ if the potential parameters satisfy the constraint $3\sqrt{a}-{b^2\over 4a}+c=0$. The search for polynomial solutions $\chi(x)=\sum \alpha_i x^i$ of (\ref{eq16}) can then be made by means of the standard techniques of series solution of second-order differential equations. However, these exact solutions can be explicitly generated by means of the asymptotic iteration method (AIM) recently introduced \cite{cs}. Actually, we should stress the usefulness of AIM as a method for determining the explicit form of the solvability constraint polynomial $P_n(E)$. AIM was first introduced \cite{cs} to solve second-order linear differential equation of the form 
\begin{equation}\label{eq17}
y''=\lambda_0(x)y'+s_0(x)y
\end{equation}
where $\lambda_0(x)\neq 0$ and $s_0(x)$ are sufficiently many times continuously differentiable. 
\vskip0.1in
\noindent\emph{Theorem 2: Given that $\lambda_0(x)\neq 0$ and $s_0(x)$ are sufficiently many times continuously differentiable, the second-order differential equation (\ref{eq17}) has the general solution
\begin{equation}\label{eq18}
u(x)=\exp\left(-\int^x\alpha dt\right)\bigg[C_2+C_1\int^x\exp\bigg(\int^t (\lambda_0(\tau)+2\alpha(\tau))d\tau\bigg)dt\bigg]
\end{equation}
if for some $n>0$,
\begin{equation}\label{eq19}
{s_n\over \lambda_n}={s_{n-1}\over \lambda_{n-1}}\equiv \alpha
\end{equation}
where 
$$\lambda_n=\lambda'_{n-1}+s_{n-1}+\lambda_0\lambda_{n-1},\quad\hbox{and}\quad s_n=s'_{n-1}+s_0\lambda_{n-1}.$$
}
\noindent The asymptotic iteration method was soon adopted \cite{cs}-\cite{ns} to investigate the solutions of eigenvalue problems of Schr\"odinger type. In such applications, one immediately faces the problem of transforming the Schr\"odinger equation (with no first derivative) into the form (\ref{eq17}). The use of asymptotic solutions of the Schr\"odinger equation under consideration is the usual approach employed to overcome this problem. It is important to mention that the asymptotic form is very crucial for the convergence of the iteration method to exact solutions. For the sextic anharmonic oscillator potential $V(x)$, or for more general even-degree polynomials $V(x)=\sum_{i=1}^{2n-1} a_i~x^{2i},~n\geq 2$, the construction based on Rainville's approach \cite{ra} provides a straightforward technique to generate a proper asymptotic form that stabilizes and accelerates the convergence of AIM. The first few iterations of (\ref{eq16}) yields polynomial expressions $\{P_n(E)\}$.  The exact eigenvalues can be computed, in turn, as the zeros of these polynomials (for convenience, we denote $P_0(E)=1, P_1(E)=1$):
\vskip0.1in
\begin{itemize}
\item $P_2(E)=2\sqrt{a}E-b$, if $12a^{3/2}-b^2+4ac=0$, $\chi_2(x)=1$.
\item $P_3(E)=2\sqrt{a}E-3b$, if $20a^{3/2}-b^2+4ac=0$, $\chi_3(x)=x$.
\item $P_4(E)=4aE^2-12\sqrt{a}bE+24a^{3/2}+3b^2+8ac$, if $28a^{3/2}-b^2+4ac=0$, 
$$\chi_4(x)=P_0(E)-{P_2(E)\over 4\sqrt{a}}x^2.$$
\item $P_5(E)=4aE^2-20\sqrt{a}bE+120a^{3/2}+15b^2+24ac$, if $36a^{3/2}-b^2+4ac=0$, 
$$\chi_5(x)=x(P_1(E)-{P_3(E)\over 12\sqrt{a}}x^2).$$
\item $P_6(E)=8a^{3/2}E^3-60abE^2+(720a^2-90\sqrt{a}b^2+112a^{3/2}c)E-552a^{3/2}b-15b^3-120abc$, if $44a^{3/2}-b^2+4ac=0$, 
$$\chi_6(x)=P_0(E)-{P_2(E)\over 4\sqrt{a}}x^2+
{P_4(E)\over 96a}x^4.$$
\item $P_7(E)=8a^{3/2}E^3-84abE^2+(1680a^2+210\sqrt{a}b^2+208a^{3/2}c)E-3480a^{3/2}b-105b^3-504abc$, if $52a^{3/2}-b^2+4ac=0$, 
$$\chi_7(x)=x\big(P_1(E)-{P_3(E)\over 12\sqrt{a}}x^2+
{P_5(E)\over 480a}x^4\big).$$
\end{itemize}
It is quite clear that the even-parity wave function solutions of sextic oscillator Hamiltonian (\ref{eq5}) satisfy
\begin{equation}\label{eq20}
\chi_{2n+2}(x)=\sum_{i=0}^n {(-1)^iP_{2i}(E)\over (2i)!(2\sqrt{a})^i}x^{2i},\quad n=0,1,2,\dots,
\end{equation}
while for the odd-parity wave function solutions we have
\begin{equation}\label{eq21}
\chi_{2n+3}(x)=\sum_{i=0}^n {(-1)^i{\mathcal P}_{2i+1}(E)\over (2i+1)!(2\sqrt{a})^i}x^{2i+1},\quad n=0,1,2,\dots.
\end{equation}
By means of the differential equation (\ref{eq6}), we see that the polynomial solvability constraints satisfy the recursion relations:
\begin{itemize}
\item For even-parity solution, with $P_0(E)=1$,
\begin{eqnarray}\label{eq22}
P_{2n+2}(E)&=&(2\sqrt{a}E-(4n+1)b)P_{2n}(E)\nonumber\\
&&+2n(2n-1)[4(4n-1)a^{3\over 2}-b^2+4ac]P_{2n-2}(E),
\end{eqnarray}
for $n=0,1,2,\dots,$ (Note that for $n=0$ the $nP_{2n-2}(E)$ term is not present) subject to the condition
\begin{equation}\label{eq23}
4(4n+3)a^{3\over 2}-b^2+4ac=0
\end{equation}

\item For odd-parity solution, with ${\mathcal P}_1(E)=1$,
\begin{eqnarray}\label{eq24}
{\mathcal P}_{2n+3}(E)&=&(2\sqrt{a}E-(4n+3)b){\mathcal P}_{2n+1}(E)\nonumber\\
&&+2n(2n+1)[4(4n+1)a^{3\over 2}-b^2+4ac]{\mathcal P}_{2n-1}(E),
\end{eqnarray}
for $n=0,1,2,\dots,$ 
(for $n=0$ the $n{\mathcal P}_{2n-1}(E)$ term is not present) subject to
\begin{equation}\label{eq25}
4(4n+5)a^{3\over 2}-b^2+4ac=0.
\end{equation}
\end{itemize}
\section{Quasi-exact solvable systems and orthogonal polynomials}
The Hamiltonian (\ref{eq5}) has the following scale transformation property
\begin{equation}\label{eq26}
H(a,b,c)={a^{1\over 4}}H(1,ba^{-{3\over 4}},ca^{-{1\over 2}})
\end{equation}
Setting $ba^{-{3\over 4}}=2t$ and $ca^{-{1\over 2}}=t^2-4J-3$, where $J$ is a nonnegative integer and $t$ is a real number, the Hamiltonian (\ref{eq5}) reads 
\begin{equation}\label{eq27}
H=-{d^2\over dx^2}+x^6+2tx^4+(t^2-4J-3)x^2.
\end{equation}
In this case, $P_n(E), n=0,1,2,\dots$ becomes
$$\cases{P_{2n}(E_{(a,b,c)})=P_{2n}(a^{1\over 4}
E_{(1,ba^{-{3\over 4}},ca^{-{1\over 2}})})\equiv P_{n}(a^{1\over 4}E) =2^na^{3n\over 4}P_n^{(t)}(E),&$n=0,1,2,\dots$\\
P_{2n-2}(E_{(a,b,c)})=P_{2n-2}(a^{1\over 4}E_{(1,ba^{-{3\over 4}},ca^{-{1\over 2}})})\equiv P_{n-1}(a^{1\over 4}E)=2^{n-1}a^{3n-3\over 4}P_{n-1}^{(t)}(E),& $n=1,2,\dots$\\
P_{2n+2}(E_{(a,b,c)})=P_{2n+2}(a^{1\over 4}E_{(1,ba^{-{3\over 4}},ca^{-{1\over 2}})})\equiv P_{n+1}(a^{1\over 4}E)=2^{n+1}a^{3n+3\over 4}P_{n+1}^{(t)}(E)& $n=0,1,2,\dots$}
$$
and the recursion relation (\ref{eq22}) now reads for $n=1,2,\dots$
\begin{equation}\label{eq28}
P_{n+1}^{(t)}(E)=(E-(4n+1)t)P_{n}^{(t)}(E)+8n(2n-1)(n-J-1)P_{n-1}^{(t)}(E),
\end{equation}
which uniquely determines, with $P_0^{(t)}(E)=1$ and $P_1^{(t)}(E)=(E-5t)P_0^{(t)}(E)$, all the polynomials $P_n^{(t)}(E), n=1,2,\dots$ in terms of $P_0^{(t)}(E)$. With these initial condition, the recursion relation (\ref{eq28}) generate a set of polynomials, the next four of which are
\begin{eqnarray*}
P_1^{(t)}(E)&=&E-t\\
P_2^{(t)}(E)&=&E^2-6tE+5t^2-8J\\
P_3^{(t)}(E)&=&E^3-15tE^2+(48(1-J)-8J+59t^2)E-3t(16-40J+15t^2)\\
P_4^{(t)}(E)&=&E^4-28tE^3+(288-176t+254t^2)E^2-4t(528-392J+203t^2)E\\
&-&48(45J-38)t^2+585t^4
\end{eqnarray*}
On  multiplying the recursion relation (\ref{eq28}) by $E^{n-1}\omega(E)$ and integrating with respect to $E$, using the fact that $P_n^{(t)}(E)$ is orthogonal to $E^k$, $k<n$, we obtain a simple, two-term recursion relation for the squared norm $\gamma_n$ of $P_n^{(t)}$ as:
\begin{eqnarray}\label{eq29}
\gamma_n&=&8n(2n-1)(J-n+1)\gamma_{n-1},
\end{eqnarray}
which is independent of $t$. The solution to this equation with $\gamma_0=1$ is
\begin{equation}\label{eq30}
\gamma_n=\prod_{k=1}^n 8k(2k-1)(J-k+1)={4^{n}~(2n)!~\Gamma(J+1)\over \Gamma(1+J-n)}.
\end{equation}The interesting factorization property \cite{bd} follows when $J$ takes nonnegative integer values. This is clear because the third term in the recursion relation (\ref{eq28}) vanishes when $n=J+1$, so that all subsequent polynomials have the common factor $P_{J+1}(E)$. To illustrate this factorization, we list in factored form the first five polynomials for the case $J=1$ 
$$P_0^{(t)}(E)=1,$$
$$P_1^{(t)}(E)=E-t,$$
$$P_2^{(t)}(E)=E^2-6tE+5t^2-8$$
$$P_3^{(t)}(E)=(E-9t)(E^2-6tE+5t^2-8)$$
$$P_4^{(t)}(E)=(E^2-22tE+117t^2+120)(E^2-6tE+5t^2-8)$$
$$P_5^{(t)}(E)=(E^3-39tE^2+(568+491t^2)E-1989t^3-6072t)(E^2-6tE+5t^2-8)$$
In general
\begin{equation}\label{eq31}
P_{n+J+1}^{(t)}(E)=Q_n^{(t)}(E)P_{J+1}^{(t)}(E)
\end{equation}
Substituting (\ref{eq31}) into (\ref{eq28}), one can obtain the recurrence relation immediately for the factor-polynomial $Q_n^{(t)}(E)$:
\begin{equation}\label{eq32}
Q_n^{(t)}(E)=(E-(4n+4J+1)t)Q_{n-1}^{(t)}(E)+8(n-1)(n+J)(2n+2J-1)Q_{n-2}^{(t)}(E)
\end{equation}
with initial condition $Q_0^{(t)}(E)=1$ so that $Q_n^{(t)}(E),n=0,1,2,\dots$ are again orthogonal polynomials. The squared norm of $Q_n^{(t)}(E)$ is given by (with $\gamma_0^Q=1$)
$$\gamma_n^Q=\prod_{k=1}^n  8k(k+J+1)(2k+2J+1).$$
Further, for the odd parity case, the recursion relation (\ref{eq24}) reads, under the scaling (\ref{eq26}),
\begin{equation}\label{eq33}
{\mathcal P}_{n+1}^{(t)}(E)=(E-(4n+3)t){\mathcal P}_{n}^{(t)}(E)+8n(2n+1)(n-J-1){\mathcal P}_{n-1}^{(t)}(E).
\end{equation}
with ${\mathcal P}_0^{(t)}(E)=1$. The first few explicit polynomials are 
\begin{eqnarray*}
{\mathcal P}_1^{(t)}(E)&=&E-3t\\
{\mathcal P}_2^{(t)}(E)&=&E^2-10tE+21t^2-24J\\
{\mathcal P}_3^{(t)}(E)&=&E^3-21tE^2+(80-104J+131t^2)E-3t(80-168J+77t^2)\\
{\mathcal P}_4^{(t)}(E)&=&E^4-36tE^3-12tE(400-312J+183t^2)+E^2(416-272J+446t^2)\\
&+&9(448(J-2)J-16(-74+77J)t^2+385t^4)
\end{eqnarray*}
Again on multiplying (\ref{eq31}) by $E^{n-1}\omega(E)$ and integrating with respect to $E$, using the fact that ${\mathcal P}_n^{(t)}(E)$ is orthogonal to $E^k$, $k<n$, we obtain a simple, two-term recursion relation for the squared norm 
${\gamma}_n^{\mathcal P}$:
\begin{eqnarray}\label{eq34}
\gamma_n^{\mathcal P}&=&8n(2n+1)(n-J-1)\gamma_{n-1},
\end{eqnarray}
The solution to this equation with $\gamma_0=1$ is
\begin{equation}\label{eq35}
\gamma_n^{\mathcal P}=\prod_{k=1}^n 8k(2k+1)(k-J-1)={4^{n}~(2n+1)!~\Gamma(J+1)\over \Gamma(J-n+1)}.
\end{equation}
It is clear that the squared norms (\ref{eq30}) and (\ref{eq35}) vanish for $n\geq J+1$, as expected. The classes of orthogonal polynomials discovered by Bender and Dunne follow directly by setting $t=0$ in (\ref{eq28}) and (\ref{eq33}).
The factorization property for the polynomials $\{{\mathcal P}_n^{(t)}(E)\}$ in the case of $J=1$ can be illustrated by means of the polynomials
$${\mathcal P}_0^{(t)}(E)=1$$
$${\mathcal P}_1^{(t)}(E)=E-3t$$
$${\mathcal P}_2^{(t)}(E)=(E-7t)(E-3t)-24$$
$${\mathcal P}_3^{(t)}(E)=((E-7t)(E-3t)-24)(E-11t)$$
$${\mathcal P}_4^{(t)}(E)=(168+(E-15t)(E-11t))((E-7t)(E-3t)-24)$$
$${\mathcal P}_5^{(t)}(E)=((E-7t)(E-3t)-24)(E^3-45tE^2+(744+659t^2)E-3135t^3-9528t)$$
In general 
$${\mathcal P}_{n+J+1}^{(t)}(E)={\mathcal Q}_n^{(t)}(E){\mathcal P}_{J+1}^{(t)}(E)$$
where ${\mathcal Q}_n(E)$ satisfy
$${\mathcal Q}_n^{(t)}(E)=(E-(4n+4J+3)t){\mathcal Q}_{n-1}^{(t)}(E)+8(n-1)(n+J)(2n+2J+1){\mathcal Q}_{n-2}^{(t)}(E),\quad n\geq 1.$$
The squared norm for these polynomials is then
$$\gamma_n^{\mathcal Q}=\prod_{k=1}^n  8k(k+J+1)(2k+2J+3)$$
Clearly, the squared norms of the polynomials $Q_n^{(t)}(E)$ and ${\mathcal Q}_n^{(t)}(E)$ do not vanish. The weight functions for all of ${P}_n^{(t)}(E)$ and ${\mathcal P}_n^{(t)}(E)$ as well as of the polynomials ${ Q}_n^{(t)}(E)$ and ${\mathcal Q}_n^{(t)}(E)$ can be computed by means of the method discussed in \cite{ak}.
\section{Conclusion}
In the field of quantum mechanics the antecedent to the concept of quasi exact solutions may perhaps be found in an early paper by Wigner \cite{wi} in 1929 in which the following simple idea is explored: first choose a wave function, and then find the corresponding potential. The form originally chosen for the wave function was the exponential of a polynomial, and it was shown that the construction worked out if the coefficients met certain conditions.  As is evidenced by the references in this paper, and indeed in the contents of the paper, many results have been discovered since that early work of Wigner. In order to test general approximation theories (so essential to applications of quantum mechanics), it is always useful to have at hand a large collection of exact solutions. This is one clear area of utility for the outcome of this work.  Of course, problems that start as questions in physics soon take on a life of their own and may later generate a repository of results that serve the field from which they originally emerged. 
\section*{Acknowledgments}
\medskip
\noindent Partial financial support of this work under Grant Nos. GP3438 and GP249507 from the 
Natural Sciences and Engineering Research Council of Canada is gratefully 
acknowledged by two of us (respectively [RLH] and [NS]).
\section*{Appendix:}
\noindent In this appendix, we search for closed-form expressions for the orthogonal polynomials, $\{P_n^{(t)}\}$ defined by (\ref{eq28}) and $\{{\mathcal P}_n^{(t)}\}$ defined by (\ref{eq33}), in terms of hypergeometric orthogonal polynomials \cite{rr}. We consider first the case of polynomials defined by the recurrence relation (\ref{eq28}). Since its squared norm (\ref{eq30}) vanishes for $n\geq 1+J$, we can take $J=n+i,~i=0,1,2,\dots$. Thus we write (\ref{eq28}) as  
\begin{equation}\label{a1}
P_{n+1}^{(t)}(E)=(E-(4n+1)t)P_{n}^{(t)}(E)-8n(2n-1)(i+1)P_{n-1}^{(t)}(E).
\end{equation}
This recurrence formula can be compared with 
\begin{equation}\label{a2}
P_{n+1}(E)=(E-(dn+f))P_{n}(E)-n(gn+h)P_{n-1}(E).
\end{equation}
studied in \cite{ch}. With $\sigma^2=4g-d^2>0$, $\delta={d\over \sigma}$, $\eta=1+{h\over g}$ and the choice $2f=\delta\eta\sigma$, the recurrence formula (\ref{a2}) becomes, for $n\geq 0$,
\begin{equation}\label{a3}
M_{n+1}(E)=[E-(2n+\eta)\delta]M_{n}(E)-(\delta^2+1)n(n+\eta-1)M_{n-1}(E)
\end{equation}
where 
\begin{equation}\label{a4}
M_{n}(E;\delta,\eta)=\bigg({2\over\sigma}\bigg)^n P_n\bigg({\sigma E\over 2}\bigg)
\end{equation}
is a Meixner polynomial of the second kind. For our case, we have $d=4t$, $f=t, g=16(i+1), h=-8(i+1)$, and, using the notation
$$\sigma^2=4g-d^2=16(4i+4-t^2),\quad{ or }\quad \sigma=4\sqrt{4i+4-t^2}$$
and
$$\delta={d \over \sigma}={t\over \sqrt{4i+4-t^2}},\quad \eta=1+{h\over g}={1\over 2},$$
the recurrence formula (\ref{a1}) becomes
\begin{equation}\label{a5}
M_{n+1}(E)=[E-{(4n+1)t\over 2\sqrt{4i+4-t^2}}]M_n(E)-{2n(2n-1)(i+1)\over \sqrt{4i+4-t^2}}M_{n-1}(E),
\end{equation}
where
\begin{eqnarray}\label{a6}
M_n(E)&=&M_n(E;{t\over \sqrt{4i+4-t^2}},{1\over 2})\\
&\equiv&
\bigg({1\over 2\sqrt{4i+4-t^2}}\bigg)^n P_n^{(t)}\bigg(2\sqrt{4i+4-t^2}E\bigg).
\end{eqnarray}
For the orthogonal polynomials $\{{\mathcal P}_n^{(t)}(E)\}$ defined by (\ref{eq33}), we again let $J=n+i,~i=0,1,2,\dots$. Thus we write (\ref{eq33}) as  
\begin{equation}\label{a7}
{\mathcal P}_{n+1}^{(t)}(E)=(E-(4n+3)t){\mathcal P}_{n}^{(t)}(E)-8n(2n+1)(i+1){\mathcal P}_{n-1}^{(t)}(E),
\end{equation}
which can be compared, again, with (\ref{a2}) for $d=4t$, $f=3t, g=16(i+1), h=8(i+1)$. In this case, we have 
$$\sigma^2=4g-d^2=16(4i+4-t^2),\quad{ or }\quad \sigma=4\sqrt{4i+4-t^2}$$
and
$$\delta={d \over \sigma}={t\over \sqrt{4i+4-t^2}},\quad \eta=1+{h\over g}={3\over 2},$$
and the recurrence formula (\ref{a7}) becomes
\begin{equation}\label{a8}
M_{n+1}(E)=[E-{(4n+3)t\over 2\sqrt{4i+4-t^2}}]M_n(E)-{2n(2n+1)(i+1)\over \sqrt{4i+4-t^2}}M_{n-1}(E),
\end{equation}
where
\begin{eqnarray}\label{a9}
M_n(E)&=&M_n(E;{t\over \sqrt{4i+4-t^2}},{3\over 2})\\
&\equiv&
\bigg({1\over 2\sqrt{4i+4-t^2}}\bigg)^n {\mathcal P}_n^{(t)}\bigg(2\sqrt{4i+4-t^2}E\bigg).
\end{eqnarray}

\end{document}